\documentclass[
a4paper,
amsmath,
amssymb
]{jpconf}
\usepackage[numbers,sort&compress,square]{natbib}
\usepackage{graphicx}
\usepackage[svgnames]{xcolor}
\usepackage{bm}

\begin{document}
\title{Spontaneous generation of fractional vortex-antivortex pairs at single edges of high-Tc superconductors}

\author{P Holmvall, T L{\"o}fwander and M Fogelstr{\"o}m}

\address{Department of Microtechnology and Nanoscience -- MC2, Chalmers University of Technology, SE-41296 G{\"o}teborg, Sweden}

\ead{holmvall@chalmers.se, tomas.lofwander@chalmers.se, mikael.fogelstrom@chalmers.se}

\begin{abstract}
Unconventional $d$-wave superconductors with pair-breaking edges are predicted to have ground states with spontaneously broken time-reversal and translational 
symmetries.
We use the quasiclassical theory of superconductivity to demonstrate that such phases can exist at any single pair-breaking facet. This implies that a greater variety of systems, 
not necessarily mesoscopic in size, should be unstable to such symmetry breaking. The density of states averaged over the facet displays a broad peak centered at zero energy, 
which is consistent with experimental findings of a broad zero-bias conductance peak with a temperature-independent width at low temperatures.
\end{abstract}

\section{\label{sec:introduction}Introduction}

It was established already in the 1990s that a number of high-temperature superconductors have an order parameter with
$d_{x^2-y^2}$ symmetry~\cite{bib:tsuei_2000}.
In such materials, quasiparticle scattering at surfaces or off defects, 
where the sign of the $d$-wave order parameter changes for incoming and outgoing scattering trajectories,
leads to the formation of Andreev bound states at zero energy~\cite{bib:hu_1994,bib:kashiwaya_tanaka_2000,bib:lofwander_shumeiko_wendin_2001}.
For an ideal specular surface with [110]-orientation, all scattering trajectories  include the sign change,
and the spectral weight of these zero-energy Andreev bound states is very large:
they form a flat band at zero energy as function of momentum parallel to the interface, $k_{\parallel}$.
Shifting these mid-gap states to finite energies can lead to lowering of the free energy.
Any mechanism providing such a shift can then lead to a phase transition
into a new ground state with an associated broken symmetry~\cite{bib:volovik_1985,bib:sigrist_1991}.
Several mechanisms have been proposed, all leading to spontaneous time-reversal symmetry breaking:
development of a subdominant superconducting component of the order parameter in a time-reversal symmetry breaking combination
with the dominant, e.g. $d_{x^2-y^2}+is$,~\cite{bib:matsumoto_1995a,bib:matsumoto_1995b,bib:fogelstrom_rainer_sauls_1997};
magnetic ordering~\cite{bib:potter_2014}; and, finally, spontaneous 
supercurrents~\cite{bib:higashitani_1997,bib:barash_kalenkov_kurkijarvi_2000,bib:lofwander_shumeiko_wendin_2000,bib:vorontsov_2009,bib:hakansson_2015,bib:nagai_ota_tanaka_2017}.
The first two scenarios require an additional coupling constant leading to an associated mean-field order parameter, while the last does not. 
Which scenario that would be realized experimentally depends on material parameters, for instance the strength of the coupling constants.
It was shown in Ref.~\cite{bib:hakansson_2015} that the transition temperature within the third scenario is very large,
of the order of 20\% of the superconducting transition temperature $T_c$.
As a consequence, the other scenarios can compete only if their corresponding coupling constants are very large,
or if the phase with spontaneous supercurrents is suppressed for one reason or another.

So far, there are several transport experiments supporting spontaneous time-reversal symmetry breaking \cite{bib:covington_1997,bib:krishana_1997,bib:dagan_2001,bib:gonnelli_2001,bib:elhalel_2007,bib:gustafsson_2013,bib:watashige_2015}.
But direct measurements of the associated supercurrents and magnetic fields remain controversial \cite{bib:neils_2002,bib:kirtley_2006}.
In our previous studies \cite{bib:hakansson_2015}, we showed that this controversy could be related to the manner in which these currents and magnetic fields appear. We found a translational and time-reversal symmetry-breaking phase, in which a staggered pattern of fractional vortex-antivortex pairs forms like a necklace along the pair-breaking surface. The symmetric proportion of vortices to antivortices effectively eliminates any net current and magnetic flux, and the small size of the vortices of a few coherence lengths makes direct observation challenging.

Vorontsov found that a phase gradient can be generated through spontaneous time-reversal symmetry breaking in thin films \cite{bib:vorontsov_2009,bib:hachiya_aoyama_ikeda_2013,bib:higashitani_miyawaki_2015,bib:miyawaki_higashitani_2015_b,bib:miyawaki_higashitani_2015_a}, caused by finite-size effects in the form of a proximity of two pair-breaking interfaces. In our previous work \cite{bib:hakansson_2015} we studied mesoscopic grains with only pair-breaking edges and found that the vortex-antivortex phase is more energetically favorable than the thin-film phase predicted by Vorontsov. In this study, we show that the vortex-antivortex phase can occur without finite-size effects. This is done by considering a system with a single pair-breaking edge.

\section{\label{sec:model}Model and methods}
We study a mesoscopic superconducting grain in vacuum and equilibrium, with a $d$-wave pairing symmetry. The sides of the system are perfectly aligned with the crystal $ab$-axes, except one facet which is misaligned by a $45^{\circ}$ rotation (see Fig.~\ref{fig:system}).
The facet gives rise to mid-gap states associated with surface pair-breaking, and has a side-length given in units of the superconducting coherence length $\xi_0 \equiv \hbar v_F / 2\pi k_BT_c$. Furthermore, a clean superconductor and a cylindrical Fermi surface is assumed.

To study this system, the quasiclassical theory of superconductivity \cite{bib:eilenberger_1968,bib:larkin_ovchinnikov_1969} is used. In this formulation, the superconducting $d$-wave order parameter $\Delta_d(\bm{R})$ depends on the anomalous Green's function (pair propagator) $f(\bm{p}_{F},\bm{R};\epsilon_n)$ through the gap equation
\begin{equation}
\label{eq:theory:gap_equation}
\Delta_d(\bm{R}) = V_dk_BT\int \frac{d\theta_{p_F}}{2\pi}\eta^{*}_{d}(\theta_{p_F})\sum_{|\epsilon_{n}|\leq\Omega_c}f(\bm{p}_{F},\bm{R};\epsilon_{n}),
\end{equation}
at spatial coordinate $\bm{R}$, quasiparticle momentum $\bm p_F$ and Matsubara energy $\epsilon_n$ (these parameters will from now on be dropped for a compact notation). Here, $\theta_{p_F}$ is the angle between the Fermi momentum and the crystal $ab$-axes, $\eta_d(\theta_{p_F})=\sqrt{2}\cos(2\theta_{p_F})$ the $d$-wave order parameter basis function, $V_d = -N_F\lambda_d$ the pair-potential, $N_F$ the normal-state density of states at the Fermi surface, $\lambda_{d}$ the pairing interaction, and $\Omega_c$ the cutoff energy.
The anomalous Green's function is the off-diagonal component of the Matsubara Green's function
\begin{equation}
\label{eq:theory:Green's_function}
\hat{g} = 
\left(
\begin{array}{cc}
{g} & {f}\\
-{\tilde{f}} & {\tilde{g}}
\end{array}\right),
\end{equation}
where \textit{hat} denotes Nambu (electron-hole) space.
The tilde operation denotes particle-hole conjugation, $\tilde{g}(\bm{p}_F,\bm{R};\epsilon_n) = g^{*}(-\bm{p}_F,\bm{R},\epsilon_n)$ [and the same for $\tilde f$].
The Green's function is obtained by solving the Eilenberger equation with the associated normalization condition
\begin{eqnarray}
\label{eq:theory:eilenberger_equation}
i\hbar\bm{v}_{F}\cdot\bm{\nabla}_{R}\hat{g} + \left[\hat{\tau}_3z - \hat{\Delta},\hat{g}\right] & = & \hat{0},\\
\label{eq:theory:normalization}
\hat{g}^2 & = & -\pi^2\hat{1},
\end{eqnarray}
where $\hat{\Delta} = i (\hat\tau_2\Re\Delta_d + \hat\tau_1\Im\Delta_d )\eta_d(\theta_{p_F})$,
and $\hat{\tau}_i$ ($i=1,2,3$) are the three Pauli matrices in Nambu space.
The Eilenberger equation and the gap equation are solved self-consistently
by the so-called Riccati technique (see for instance Ref.~\cite{bib:eschrig_2009}). After self-consistency has been achieved, we compute observables such as the current density
\begin{equation}
\label{eq:theory:current_density}
\bm{j}(\bm R) = 4\pi eN_Fk_BT \int \frac{d\theta_{p_F}}{2\pi}  \bm{v}_F(\bm{p}_F) \sum_{\epsilon_n} g(\bm p_F,\bm R;\epsilon_n).
\end{equation}
The magnetic flux density induced by the current density is calculated through Maxwells equations and Amp{\`e}re's circuit law.

\section{\label{sec:results}Results and discussion}
\begin{figure}[b]
\includegraphics[width=1.0\textwidth]{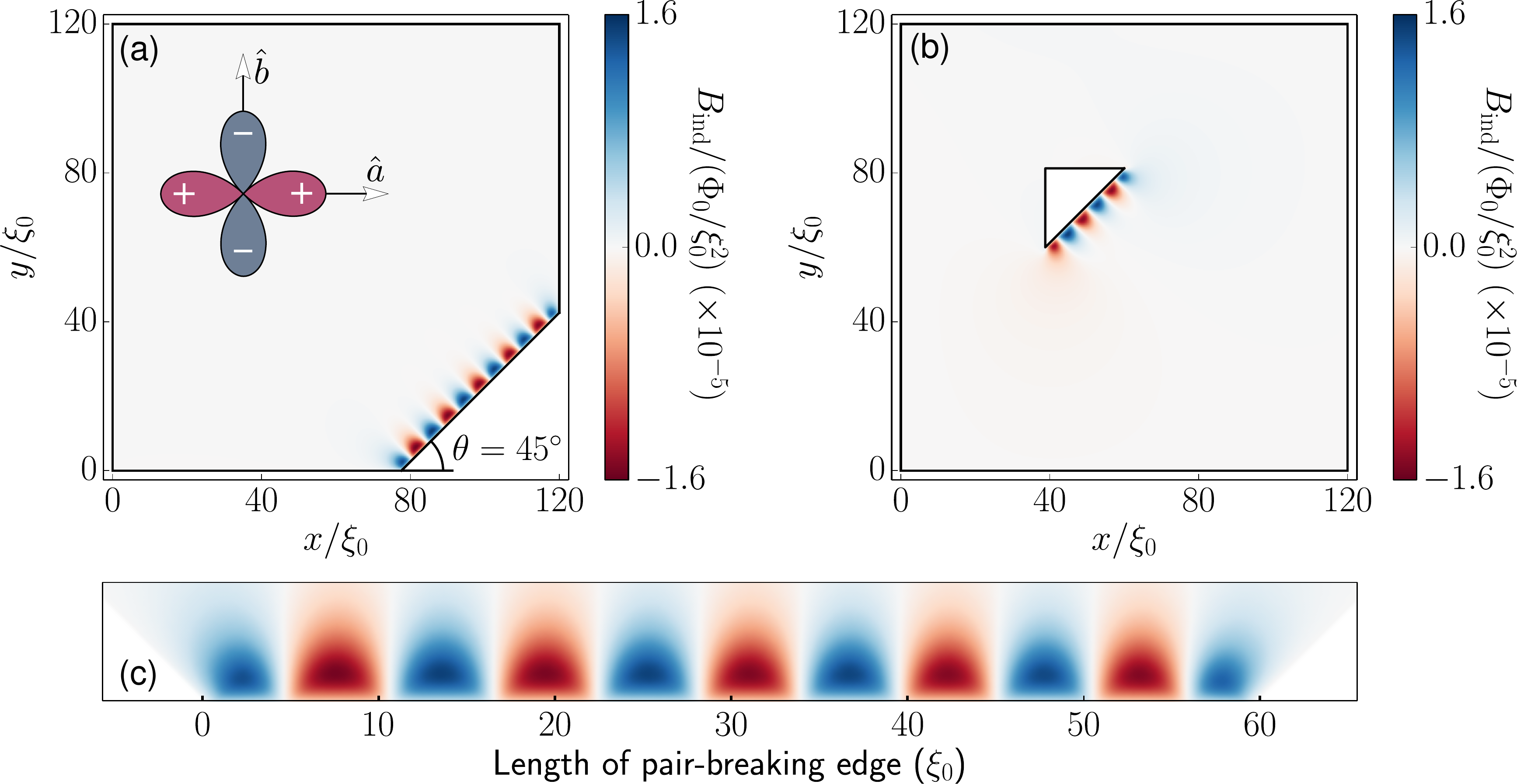}
\caption{\label{fig:system} (Color online)
(a) A $d$-wave superconducting grain at temperature $T=0.1T_c$ with a spontaneously induced magnetic flux density, due to spontaneous generation of fractional vortices and antivortices. The latter breaks time-reversal and translational (along the facet) symmetries, and is linked to an energetically favorable Doppler shift of mid-gap states to finite energies. These mid-gap states are formed through pair-breaking along the diagonal grain facet, which is rotated $45^{\circ}$ relative to the crystal $ab$-axes. All other grain edges are perfectly aligned with the crystal axes, as indicated by the graphics. In panel (b), a triangular portion of a square superconductor has been cut away, such that the pair-breaking facet is surrounded by bulk superconductivity. Panel (c) is a magnification of the pair-breaking facet in panel (a).}
\end{figure}

\begin{figure}[b]
\includegraphics[width=1.0\textwidth]{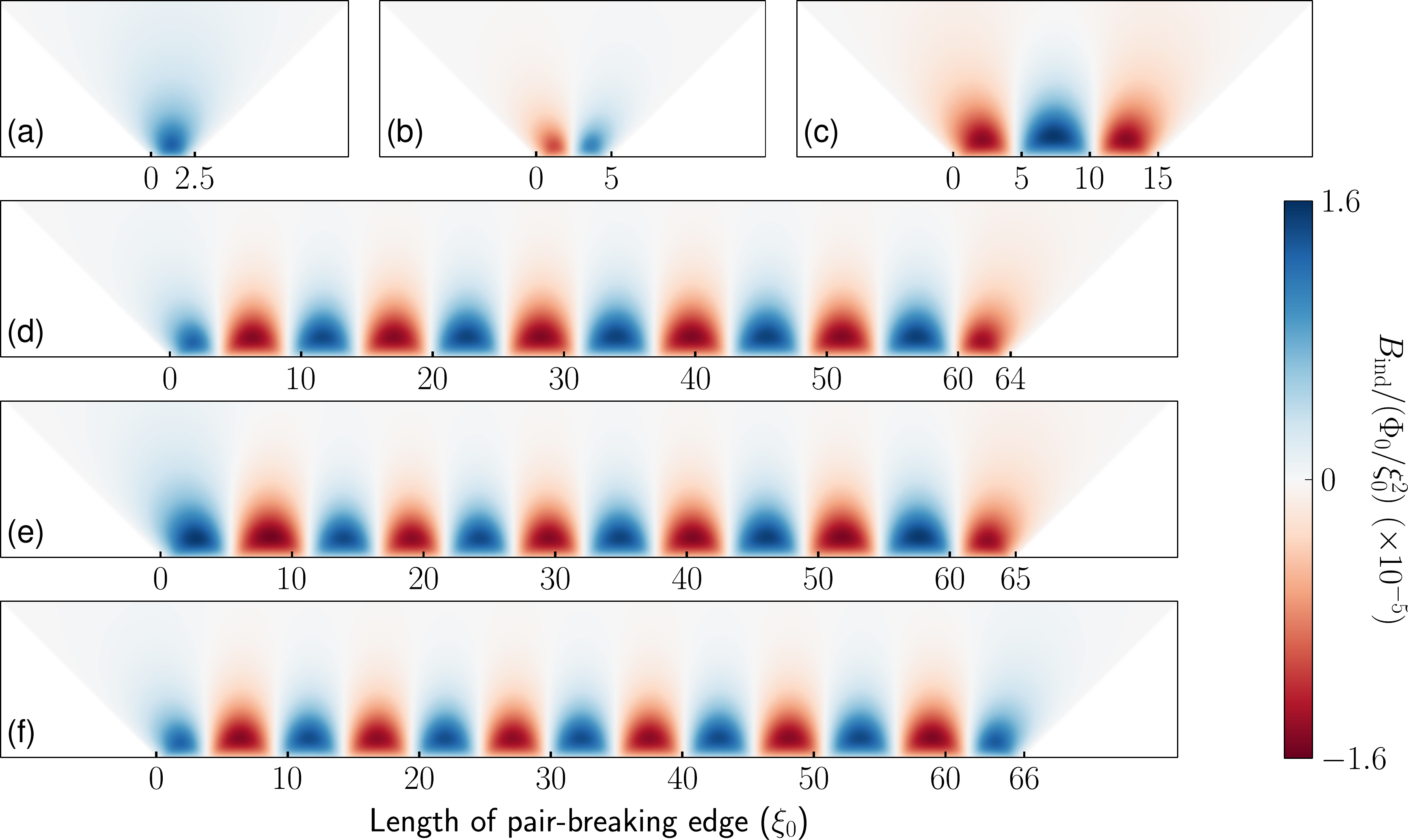}
\caption{\label{fig:collage} (Color online) Magnetic flux density due to spontaneous fractional vortices along a pair-breaking facet, where the length of the facet varies from panel (a) to (f). In each panel, the system is a square grain with a side-length of $120 \xi_0$, with one of the corners cut off at a $45^{\circ}$ angle to generate the pair-breaking facet, as illustrated in Fig.~\ref{fig:system} (a). Due to finite-size effects, there might be an unequal amount of vortices and antivortices, although the total flux still adds to zero. The only exception is when the facet is smaller than the typical vortex size ($\sim 5\xi_0$) as in panel (a), at which point there is a single vortex and a net flux. The temperature is $T=0.1T_c$.}
\end{figure}

Figure~\ref{fig:system} shows the induced magnetic flux density for two different superconducting grains that both have a single pair-breaking facet.
The flux is generated by the fractional vortex-antivortex phase, and the pair-breaking facet is formed by cutting away either a triangular corner or a triangular section in the middle of a square grain, as shown in Figs.~\ref{fig:system} (a) and (b), respectively. Thus, in the latter case, the pair-breaking facet is completely surrounded by bulk superconductivity. The fact that the phase persists in these two systems clearly illustrates a contrast to the Vorontsov phase \cite{bib:vorontsov_2009}, which relies on the proximity of two pair-breaking edges. Figure~\ref{fig:system} (c) shows a magnification of the pair-breaking facet in Figure~\ref{fig:system} (a). As shown, there might be an unequal number of vortices and antivortices for certain sizes, although the flux density sums to zero. This is illustrated further in Fig.~\ref{fig:collage}, where we vary the length of the pair-breaking facet in corner-cut systems.
Each panel shows the induced flux along the pair-breaking facet in a square grain of side-length $120 \xi_0$. The length of the facet varies from $2.5 \xi_0$ in panel (a), to $66 \xi_0$ in panel (f). There are two relevant regimes; one when the length of the facet is comparable to the fractional vortex size ($\sim 5 \xi_0$), and another when it is much larger. In the latter case, the fractional vortices have a fairly constant diameter of $5 \xi_0$, except the corner, or  (outermost, vortices which are generally smaller. Lengthening the facet increases the size of the corner vortices, until they reach the same size as the central vortices, and new corner vortices are formed. Therefore, there might be an unequal number of vortices and antivortices for certain sizes. The flux density sums to zero, however, thanks to the corner vortices being much smaller. This again illustrates the fractionality of the vortices. The most striking feature, however, is that the phase survives even as the facet becomes smaller than $5 \xi_0$, yielding a system with a single fractional vortex and a clear net flux. The system obviously finds it more favorable to shift the mid-gap states at the expense of having a net flux. Thus, the system with a single pair-breaking facet seems to lack a critical minimum size, in contrast to both the thin-film geometry \cite{bib:vorontsov_2009}, and the mesoscopic grain where all sides are pair-breaking \cite{bib:hakansson_2015}.

\begin{figure}[t]
\includegraphics[width=1.0\textwidth]{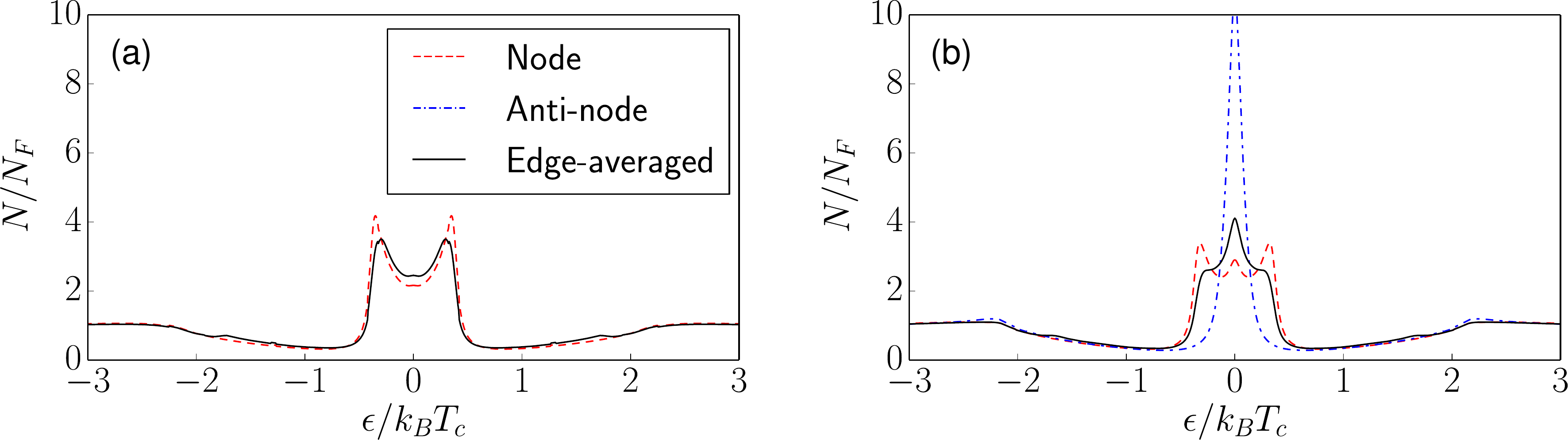}
\caption{\label{fig:ldos} (Color online) Density of states as a function of energy at the pair-breaking facet, evaluated in the middle of a vortex current (dashed line) and between vortices (dot-dashed line). The solid line is the facet-averaged density of states. Panels (a)--(b) correspond to the systems in Figs.~\ref{fig:collage} (a)--(b), respectively. The rest of the systems have an identical DOS as in panel (b).}
\end{figure}
Finally, Figs.~\ref{fig:ldos} (a)--(b) show the density of states (DOS) along the facet for the systems in Figs.~\ref{fig:collage} (a)--(b), respectively.
All other systems have an identical DOS as in panel (b). The solid lines represent the facet-averaged DOS, the dashed lines the local DOS at a node (vortex) and the dot-dashed lines the local DOS at an anti-node (between vortices). System (a) has a single vortex, resulting in a fully split peak. All other systems show a wide peak in the facet-averaged DOS. This result would be observable in a tunneling experiment as a conductance peak centered at zero energy with a rather large width, that at low temperatures is temperature independent. Only for system (a), or with a very local probe (point contact with a diameter smaller than the superconducting coherence length) would a split conductance peak be observable.

\section{\label{sec:conclusions}Conclusions}
We have used the quasiclassical theory of superconductivity to study a phase that spontaneously breaks translational and time-reversal symmetries at pair-breaking edges, in unconventional $d$-wave superconductors. Similar phases have been suggested by theory for quite some time, but up until now, have relied on finite-size effects and the proximity of two such pair-breaking edges. We have shown that such finite-size effects are not necessary for such a phase to exist, and that there is no clear critical size below which the phase disappears. This implies that any system with pair-breaking edges should be unstable to the formation of fractional vortices. Therefore, the phase should be present at a greater variety of systems than previously proposed, and lead to a broadening of zero-bias conductance peaks.

\ack
We thank the Swedish Research Council (VR) for financial support. It is a pleasure to thank Mikael H\r{a}kansson, Niclas Wennerdal and Anton Vorontsov for valuable discussions.


\providecommand{\newblock}{}

\end{document}